# Atomic Layer deposition of 2D and 3D standards for quantitative synchrotron-based composition and structural analysis methods

Nicholas G. Becker, [1, 2] Anna Butterworth, [3] Andrey Sokolov, [2] Muriel Salome, [4] Steven Sutton, [5] De Andrade Vincent, [6] Andrew Westphal, [3] and Thomas Proslier [7]

1, Physics Department, Illinois Institute of Technology, Chicago, Illinois 60616, USA

2, Material Science Division, Argonne National Laboratory, Lemont, Illinois 60439, USA

3, Lawrence Berkley National Laboratory, Berkley, California 94720, USA,

4, European Synchrotron Radiation Facility, Grenoble 38000, France

5, GSECARS, University of Chicago, Argonne, Illinois 60439, USA

6, X-Ray Science Department, Advanced Photon Source, Argonne National Laboratory, Lemont 60439, USA

7, Commissariat de l'énergie atomique, Centre Saclay, Gif-sur-Yvette 91191, France

E-mail: thomas.proslier@cea.fr

May 23, 2017

Abstract

The use of Standard Reference Materials (SRM) from the National Institute of Standards and Technology (NIST) for quantitative analysis of chemical composition using Synchrotron based X-Ray Florescence (SR-XRF) and Scanning Transmission X-Ray Microscopy (STXM) is common.

These standards however can suffer from inhomogeneity in chemical composition and thickness and often require further calculations, based on sample mounting and detector geometry, to obtain quantitative results. These inhomogeneities negatively impact the reproducibility of the measurements and the quantitative measure itself. Atomic Layer Deposition (ALD) is an inexpensive, scalable deposition technique known for producing uniform, conformal films of a wide range of compounds on nearly any substrate material. These traits make it an ideal deposition method for producing films to replace the NIST standards and create SRM on a wide range of relevant 2D and 3D substrates. Utilizing Rutherford Backscattering, X-ray Reflectivity, Quartz crystal microbalance, STXM, and SR-XRF we show that ALD is capable of producing films that are homogenous over scales ranging from 100's µm to nms.

## Introduction

Synchrotron Based X-Ray Florescence (SR-XRF) and Scanning Transmission X-Ray Microscopy (STXM) are two powerful techniques that have chemical speciation resolution capabilities from hundreds of microns to tens of nanometers.[1{6] These techniques allow for the speciation and quantification of major and trace elements, in vacuum and ambient environments, with detection limits on the order of parts per million. STXM can be viewed as a combination of soft X-ray Absorption Spectroscopy and sub-micron microscopy[3] with chemical spatial resolution down to tens of nanometers.[7] This technique is used in fields ranging from characterization of basaltic glass[1] to in situ investigation of organic field effect transistors.[4] SR-XRF can operate at higher energies, with spatial resolution on the order of a few microns down to hundreds of nanometers. Ergo, SR-XRF is utilized in a variety of fields including the visualization of lost paintings,[6] the quantification of trace elements in individual protist cells,[8] the study of metal homeostasis in plants,[9] and the distribution patterns of trace elements, such as Arsenic in rice.[10]

These techniques however suffer from the inability to quantitatively measure constituents without the use of standards of some known areal density, usually expressed in g/cm². The majority of research done using these techniques utilizes either NIST produced thin lm 2D standards, or solution based standards that are then diluted and measured, either in solution or after drying on an appropriate substrate. In either case the assumption is that the standard is homogenous on a large and small scale. In addition, the rapid development of nanoscale 3D imaging TXM techniques and powerful reconstruction algorithms for nanomaterial sciences in the field of energy storage, microelectronics, light harvesting materials calls for the development of new 3D standards of known shape and composition.

Regardless of the standard, algorithms must be used to correct for the difference in absorption between reference and sample using a fundamental parameters approach. Further complications arise when standard reference material (SRM) does not exist for a specific element of interest. In these cases the use of an SRM with elements that have atomic numbers above and below the element of interest, along with software packages to extrapolate the sensitivity for the element of interest, allows for quantitative determinations.[9]

It is commonly accepted that using these standards requires precise alignment on the previously analyzed spot, otherwise the quantitative analysis will be incorrect. Consideration must also be given to the substrate that will be utilized during measurement, whether it is a sample collection lm, or a membrane support, unknown quantities of impurities can influence the quantification. Further complications are introduced by the inability to measure the incident intensity of the x-ray beam without removing the standard. These deficiencies can be overcome by the production of highly homogenous standards made by ALD on perforated or holey (sic) Transmission Electron Microscopy (TEM) windows for 2D standards, or on 3D structures made by high resolution 3D printing techniques.

Atomic Layer Deposition (ALD) is a self-limiting, vapor deposition technique that pulses consecutive doses of chemical reagents to deposit thin films in a layer-by-layer fashion. It is characterized by large area uniformity, conformality over arbitrarily complex-shaped samples and atomic scale thickness control.[11,13] The library of materials that can be deposited via ALD is extensive and encompasses the majority of the non-radioactive portion of the periodic table.[14] These unique capabilities have been used in a large variety of applications for which atomic scale thickness and composition control on industry production levels (over hundreds of square meters) is crucial.[15,22] Here we report on the measurement of thin ALD lms uniformity deposited on thin, at, TEM membranes and 3D cubes made by 3D printing with the goal of replacing the conventional 2D SR-XRF and STXM standards as well as creating new 3D standards in anticipation of future advances in 3D imaging synchrotron techniques.

## Experimental

ALD films were grown in a commercial UltraTech (formerly Cambridge NanoTech) Savannah 100 or a custom-built ow reactor described elsewhere.[16] The UltraTech system has been modified to accept bubbler type precursor cylinders to aid in the deposition of compounds using low vapor pressure precursors. Chemicals were obtained from Strem Chemicals and Sigma Aldrich and used as received. Unless otherwise noted precursors were held in stainless steel cylinders. All compounds were prepared using previously disclosed ALD chemistries, listed in Table 1.

### 2D standards

The 50 nm thick $Si_3N_4$ TEM windows were obtained from Norcada (part# NH050A3 and NT050C) (Fig.1a) and b)) and used as received. TEM windows were held in a specially designed stainless steel holder to keep four TEM windows in place during the cycling from the ALD deposition pressure of 1 Torr to atmosphere and shown in Fig.1 c) and d).

The pumping and venting of the ALD chamber have to be done gradually and slowly to avoid breaking the thin TEM windows. Films were simultaneously deposited on 400 μm thick Si (001) and sometimes $Si_3N_4$ coupons which were cleaned via sonication in acetone, isopropanol, and then a rinse in deionized water. Prior to depositing targeted standard films, the substrates were coated with a thin (20 cycles) $Al_2O_3$ layer to alleviate thickness uncertainty that could arise from poor nucleation on inert $Si_3N_4$.

Table 1: Precursor chemistries, dose time, purge time, and deposition temperature for all compounds measured. These parameters are for 2D standards unless otherwise noted.

| Compound | Metal Precursor (dose(s)/purge(s)) | Oxygen/Nitrogen Source (dose(s)/purge(s)) | Deposition Temperature (°C) | Reference |
|---|---|---|---|---|
| $Fe_2O_3$[a] | $FeCl_3$[b] (2/60) | $H_2O$ (0.1/60) | 250 | 23 |
| $Fe_2O_3$[c] | $FeCl_3$[b] (3/20) | $H_2O$ (2.5/20) | 300 | 23 |
| $Y_2O_3$[c] | $Y(Cp)_3$[b] (7/40) | $H_2O$ (2/20) | 250 | 22 |
| $TiO_2$[c] | $TiCl_4$ (2/20) | $H_2O$ (2/20) | 250 | 24 |
| $Al_2O_3$[a] | TMA (0.1/20) | $H_2O$ (0.1/20) | 200 | 20 |
| $Al_2O_3$[c,d] | TMA (1/10) | $H_2O$ (1/10) | 165 | 20 |
| ZnO[a] | DEZ (.2/60) | $H_2O$ (.1/60) | 140 | 20 |
| ZnO[c,d] | DEZ (1/10) | $H_2O$ (1/10) | 165 | 20 |
| MgO[c] | $Mg(Cp)_2$[b] (1.5/25) | $H_2O$ (1/20) | 250 | 25 |
| $Er_2O_3$[c] | $Er(MeCp)_3$[b] (2/25) | $H_2O$ (1/15) | 250 | 26 |
| MoN[c] | $MoCl_5$[b] (1.5/15) | $NH_3$ (1/15) | 450 | 27 |

[a] UltraTech Reactor
[b] Held in bubbler style cylinders
[c] Custom Built Reactor
[d] 3D standard

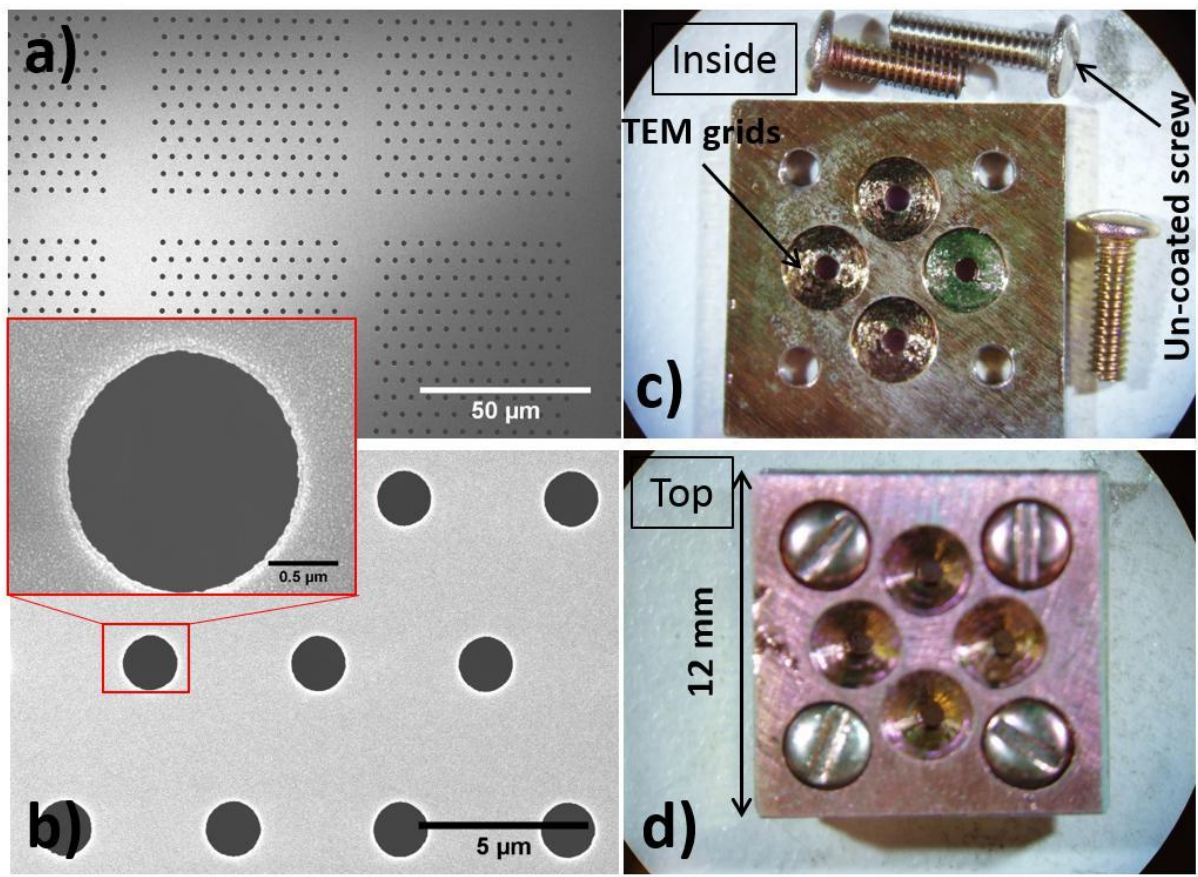

Figure 1: a) and b) SEM pictures of a half holey TEM Si$_3$N$_4$ window coated with 300 cycles of ZnO. c) and d) the TEM windows holder used for ALD depositions. Note the uniformity of the window holder color due to the past depositions and indicative of a homogeneous coating even on the screws' threads (an uncoated screw is shown for comparison).

STXM measurements were done at the Advanced Light Source (ALS), beamline 11.0.2 where the L-edges of Iron and Zinc, and the K-edges of Aluminum were investigated. SR- XRF measurements of Yttrium, Iron, and Zinc were carried out at both the ALS, beamlines 5.3.2.1 and 10.3.2, and the Advanced Photon Source (APS), beamline 13ID, where Titanium SR-XRF was also collected.

The areal (or columnar) density is determined by measuring the optical density across an edge jump. In all cases the measured areal densities (µg/cm$^2$) are expected to be double that deposited on a silicon witness piece due to the conformal nature of ALD where both sides of the TEM windows are coated during a deposition. For a given sample the optical density, OD, at a given position is

$$OD = \mu \cdot \rho \cdot t = -ln\left(\frac{I}{I_0}\right) \qquad (1)$$

where µ is the absorption cross section, ρ is the density in g/cm$^3$, t is the thickness, *I* is the measured transmitted intensity, and *I$_0$* is the incident intensity. The absorption cross section changes with incident energy and by fitting the change for any given edge jump the areal density can be obtained. ALS measurements were analyzed using the aXis2000 software package. The mass absorption coefficient, µ, is obtained from reference[28] and is also available within the aXis2000 software. Areal density is computed using a 2-component t of an OD spectrum with the computed mass absorption coefficient spectra, µ, for: 1) the element of interest and 2) the major composition of the sample contributing to the pre-edge. This fitting procedure removes the pre-edge portion of the spectrum and allows for the fitting of the post edge, with extrapolation back to the edge jump, such that the magnitude of the optical density can be calculated. The APS measurements were performed in a similar manner, but instead of using calculated absorption coefficients, a thin film reference standard from NIST (NIST SRM-1833) was utilized, making the measurement as free of assumptions as possible. In both cases the areal density, t, was computed.

The NIST standards are tested using multiple analytical techniques (atomic absorption spectrometry, inductively coupled plasma emission spectrometry, neutron activation analysis, direct current plasma emission spectrometry, and isotope dilution thermal ionization mass spectrometry), and a percent uncertainty is provided with the certificate of analysis.

The error associated with the aXis2000 fitting procedure arises from both the t itself, with a quoted 1σ error, and the tabulated values of μ, ± 10%. The precision of the measurements depends on the signal to noise ratio of the spectrum and varies widely (5% to 100%) depending on thickness, the particular edge, and the abundance of the element. To insure the greatest measurement precision the ALD films were deposited with an "optimal" OD value between 0.1 and 2.0, for which the count detection calibrations are linear.

Rutherford Backscattering (RBS) measurements The XRF measurements of Mg (K edge), Er (M edge) and Mo (L edge) were done at ESRF beamline 21 at an energy of 2.7 keV, the beam was focused with Kirkpatrick-Baez mirrors down to 0.6 m x 0.84 m. The detector is a Bruker Silicon drift diode collimated to 80 mm². A 100 s static spectrum and a 2D map with 1 m steps over 50x50 m² with 300 ms integration time per point were acquired for each samples. At ESRF, the spectra and maps were fitted with PyMCA? to deconvolve the elemental maps. The respective elements mass fraction were calculated using a NIST standard and considering a 50 nm $Si_3N_4$ matrix then converted to g/cm² by multiplying the mass fractions by the matrix density (3.44 g/cm³) for $Si_3N_4$ and thickness (50 nm). The obtained results were then normalized to the I0 map and multiplied by iodet value corresponding to the photon flux used in PyMCA configuration file.

Several methods were used to measure independently the elements areal density on at Si or $Si_3N_4$ coupons grown at the same time as the TEM windows. Rutherford Backscattering (RBS) measurements were performed on silicon witness pieces by Evans Analytical Group using a 2.275MeV He$^{++}$ ion beam, normal to the sample surface and a backscatter detector oriented at 160° or at 110°. Values for the areal density of the metal ions, $A_d$, were then calculated from the atomic areal density of the oxide, $_{RBS}$, the atomic mass, $m_a$, Avogadro's number, $N_A$, and the reported percentage of the ion of interest, x, using equation 2.

$$A_d = x \cdot \frac{\rho_{RBS} \cdot m_a}{N_A} \qquad (2)$$

X-ray Reflectivity (XRR) is another method used to extract the elemental areal density with equation 3:

$$A_d = \frac{\rho_{XRR} \cdot t \cdot m_a}{m_b + \eta \cdot m_a} \qquad (3)$$

where $\rho_{XRR}$ is the film density in g/cm$^3$, t is the lm thickness and $m_a$ and $m_b$ the atomic mass of elements a and b. $\eta$ is the element a stoichiometry normalized to b; for instance for Y$_2$O$_3$ = Y0$\eta$ with $\eta$ = 1.5. XRR measurements were performed on a Philips X'Pert ProMRD diffractometer using Cu Kα radiation ($\lambda$ = 1.5418 Å) and operated at 30 kV/30 mA. The incident X-ray beam was conditioned by a 60 mm graded parabolic W/Si mirror with a 0.8° acceptance angle and a 1/32 divergence slit. The reflected beam was collected with a PW3011/20 sealed proportional point detector positioned behind a 0.27° parallel plate collimator and a pyrolytic graphite monochromator.

Finally, in-situ Quartz Crystal Microbalance (QCM) was used to monitor the ALD lm growth. QCM measures in real time the deposited alloy areal density $A_{dQCM}$ and the element specific areal density can be obtained by:

$$A_d = \frac{A_{dQCM} \cdot m_a}{m_b + \eta \cdot m_a} \qquad (4)$$

Due to the design of the custom built ALD ow reactor, the QCM measurements were not done simultaneously with the thin lm standard growth but under the same conditions (temperature, Pressure, dose and purge times).

## 3D standards

The 3D standards were made using the 3D printer model Photonic Professional GT from Nanoscribe@ that enables 3D micro and nanofabrication via two-photon polymerization. The shape of a cube with the corresponding edges represented in Fig.2-b), was chosen as a reference structure that provides different line directions with respect to the X-ray beam (represented by an arrow in Fig.2-b)) in order to test the reconstruction algorithm. This structure, 5.7 m side size, was 3D printed at the apex on a Tungsten tip with a radius of curvature of 2 m (Fig.2-a)). The tip was laid down horizontally on a glass substrates, maintained immobile with a tape and immersed in the photoresist IP-L 780. The W tip was aligned carefully with respect to the laser beam (780 nm, same direction as the X-ray arrow) focal point at low power prior to starting the writing process with the parameters: 25% of the maximal laser power and a power scaling of 0.9. The tip was moved by a piezo stage with respect to a stationary focused laser beam at an optimal writing speed range between 20 to 30 m/s.

The shape of the laser focal point, so called vertex, is an ellipsoid with a nominal minimal size of 1 m along the laser beam direction and 0.3 m in the perpendicular plan. The vertex shape is responsible for the cube edges asymmetry and atten aspect. As can be seen in Fig.2-b), we choose to start the writing process before the tip apex in order to anchor the cube on the tip and provide better stability during future handling. The structures were then developed by dipping the tips into PGMEA for 30 minutes then clean with isopropanol and dry in air. After about 5 optimization attempts, 3 cubes were successfully printed among which two are represented in Fig.2-c-d). The measured cube side and edge dimensions are 5.6 µm and 0.5x1.5 µm respectively, in close agreement with the nominal design speci cations. The tips were then inserted into the custom-built ALD chamber and a multilayer com-posed of $Al_2O_3$ and ZnO was deposited at 165°C with the targeted thicknesses, t, and deposition parameters summarized in Table 2. The actual layer thicknesses and the corresponding ALD growth rates (GR) were measured by X-ray reflectivity on witness Si coupons.

The TXM measurements were made at the Advanced Photon Source at beamline 32-ID-C.[29] The dataset consists in 1501 projections with 500 ms exposure acquired at 8 keV in absorption mode. The X-ray objective lens of the microscope was a 180 m large Fresnel zone plate with an outermost zone width of 60 nm and a thickness of 1.4 m for an e ciency of 18%. The 2D optical resolution is around 60 nm. Reconstruction were performed using thetoolkit Tomopy[30,31] calling a ring artifact removal algorithm[32] and the sirt-fbp reconstruction algorithm.[33-35] The reconstructed volume as a voxel size of 28 nm.

Table 2: multilayer ALD parameters deposited on the 3D cube.

| Compound | ALD cycles | Targeted t (nm) | Measured t (nm) | Measured GR (A/cy) |
|---|---|---|---|---|
| $Al_2O_3$ | 50 | 5 | 6  0.03 | 1.2 |
| ZnO | 28 | 5 | 4.2 0.0 | 1.5 |
| $Al_2O_3$ | 100 | 10 | 12  0.03 | 1.2 |
| ZnO | 55 | 10 | 8.7  0.03 | 1.58 |
| $Al_2O_3$ | 160 | 16 | 19  0.13 | 1.18 |
| ZnO | 111 | 20 | 18  0.16 | 1.62 |
| $Al_2O_3$ | 400 | 40 | 41.5 0.16 | 1.04 |
| ZnO | 222 | 40 | 39.6  0.2 | 1.77 |
| $Al_2O_3$ | 800 | 80 | 83.6 0.25 | 1.04 |
| ZnO | 444 | 80 | 79.6 0.26 | 1.77 |

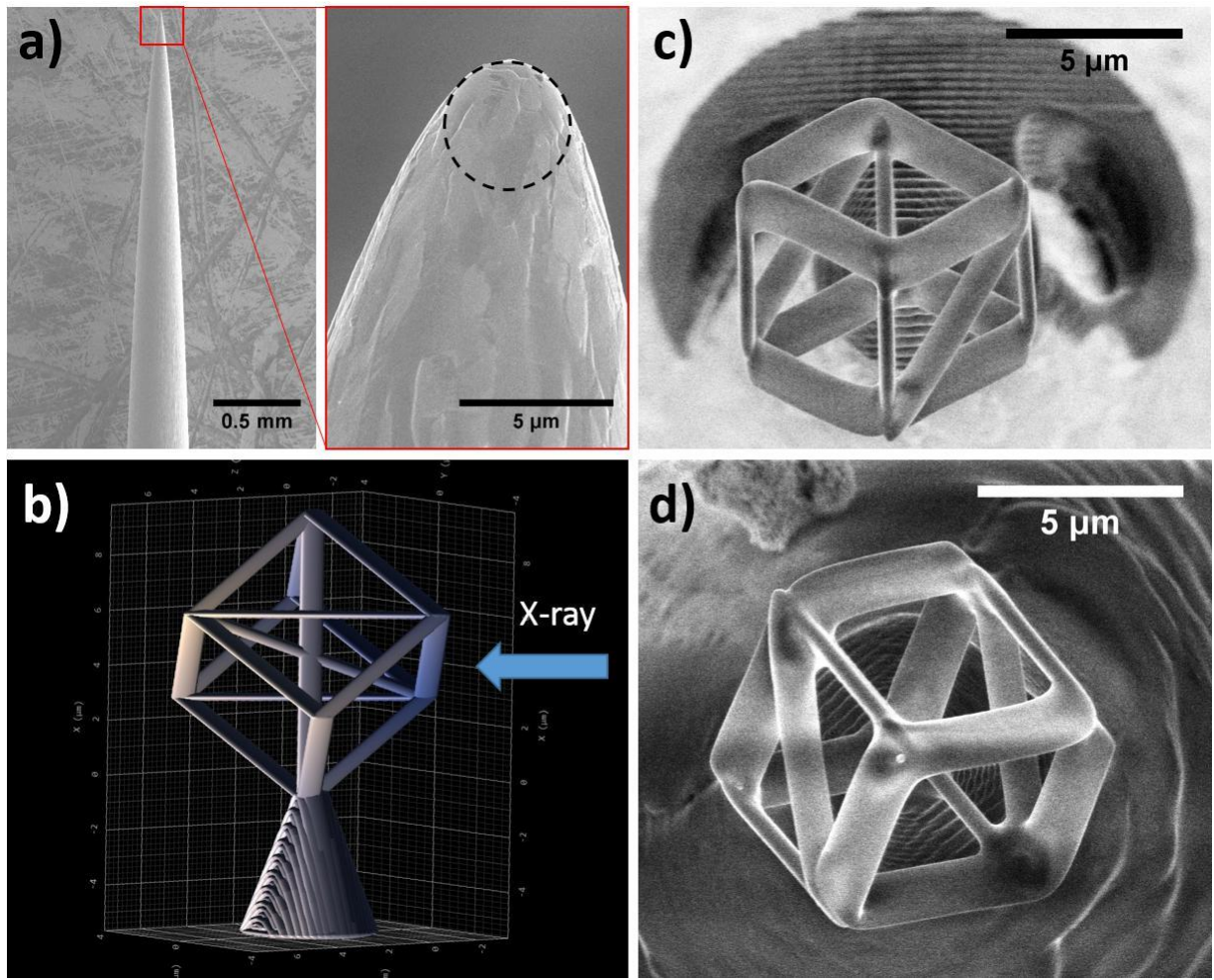

Figure 2: a) SEM image of a W tip with a curvature radius of 2 m. b) representation of 3D cube structure from the nanoscribe@ laser writing software. c) and d) SEM images of two cubes viewed from the top after the laser writer and development process.

# Results and Discussion

## 2D standards

Two types of standards were constructed, single compound films and multilayer films. When constructing multilayer standards some consideration as to order must be given to eliminate possible detrimental surface reactions.[20] In Table 3 compounds denoted with and $^+$ are portions of the same multilayer films. ALD enable the synthesis of multilayer structures with various growth sequences; either the films are grown sequentially as it was done in as follow: Substrate/ $Al_2O_3$ - 20 cy / $TiO_2$ - 250 cy / $Fe_2O_3$ - 200 cy / $Y_2O_3$ - 500 cy/ ZnO -150 cy, or the various alloys can be diluted into a matrix such as in the multilayer $^+$: Substrate/ $Al_2O_3$ - 20 cy/ 60 x {20 cy MgO + 1 cy $Er_2O_3$}. Although the sequence itself does not matter for a targeted element areal density (as long as the growth processes are well controlled), the diluted approach highlight the minimal deposition possible by ALD, i.e. one ALD cycle, that varies typically between 10 to 100 ng/cm$^2$/cy depending on the material synthesized. This quantum of areal density deposited, listed for each cation element in Table 3, set the sensitivity limit for very low concentration standards made with one ALD cycle into an arbitrary matrix.

The results from the STXM/SR-XRF measurements on TEM windows and those measured by RBS, XRR and QCM on silicon witness pieces are given in Table 3 and summarized in figure 3. The fitting and measurement uncertainty of the RBS, XRR and QCM techniques are taken into account in the errors listed. These three techniques give very consistent results with an average uncertainty of 1.5 0:8%, which is about one order of magnitude smaller than the STXM/XRF measurements and analysis errors using the NIST standards: 10 2:4%. All measured areal densities values on silicon witness pieces are in good agreement with the STM/XRF ones on TEM windows. Some systematic difference however exist which could be attributed to several factors: variation in the NIST calibration standards and fitting procedures used in STXM/XRF analysis, or non-conformal coating by the deposition method. The latter factor has been seen previously[12,36] when the deposition temperature is outside the ALD regime and is discussed later in more detail. Within the former factor, it is important to mention that the possibility of small scale variation outside of the percent uncertainty listed cannot be ignored. This is particularly important in STXM or micro SR-XRF where the measurement can be very localized. STXM and SR-XRF measurements were carried out on a TEM membrane and a Silico n witness piece respectively, both coated simultaneously with $Fe_2O_3$. The STXM measurement yields an areal density value of 14.2 g/cm$^2$, exactly double the one obtained with the SR-XRF measurement 7.1 g/cm$^2$. This is to be expected as ALD coats conformaly both sides of the thin TEM windows and STXM probes the full structure thickness: (thin films coating)/ TEM window (50 nm)/ (thin films coating) whereas SR-XRF probes only one side of the ALD coated 400 m thick witness silicon coupon.

This result comforts the fact that ALD coating inhomogeneity cannot be the source of discrepancy mentioned earlier and further emphasize the need for new, reliable standards in STXM/XRF quantitative analysis.

For this work all depositions were carried out at temperatures within this self-limiting window, which should eliminate the possibility of parasitic chemical vapor deposition (CVD) and the inhomogeneous coating associated during the ALD process. The substrate heating due to the exothermic nature of ALD reactions however may play a role. A simple formula for the change in temperature as a function of substrate, thickness, t, density, ρ , change in enthalpy, ΔH, number of reaction sites, σ, and the specific heat of the substrate, c, is given in equation 5;

$$\Delta T = \frac{2 \cdot \sigma \cdot \Delta H}{c \cdot N_A \cdot \rho \cdot t} \qquad (5)$$

Table 3: Areal density values, $A_d$ in g/cm$^2$, measured via STXM/XRF on TEM windows and compared to those measured on silicon witness coupons with RBS, XRR and QCM and extracted using equations 2, 3, 4. The minimal areal density deposited per ALD cycle for each cation, so called Quantum ALD $A_d$, is in ng/cm$^2$/cy and correspond to the average of the RBS, XRR and QCM measurements divided by the number of ALD cycles.

| Compound | Cation | $A_d$ STXM/XRF | $A_d$ RBS | $A_d$ XRR | $A_d$ QCM | Quantum ALD $A_d$ |
|---|---|---|---|---|---|---|
| $Fe_2O_3$ | Fe | 14.2  1.42[a] | 15.85  0.40 | 15.6  0.8 | 16.1  0.3 | 26.4  0.8 |
| $Fe_2O_3$ * | Fe | 10.84  0.35[b] | 10.69  0.12 | 10.77  0.06 | 10.4  0.18 | 26.5  0.3 |
| $Y_2O_3$ * | Y | 39.09  3.91[b] | 34.41  0.97 | 35.6  1.1 | 36.8  0.8 | 35.6  1.4 |
| $TiO_2$ * | Ti | 8.61  1.21[b] | 8.07  0.24 | 8.25  0.1 | 8.1  0.15 | 16.4  0.3 |
| ZnO* | Zn | 25.05  2.08[b] | 24.72  0.57 | 24.81  0.07 | 24.6  0.08 | 83.2  0.8 |
| ZnO | Zn | 55.3  5.53[a] | 52.2  0.5 | 52.8  0.4 | 53.1  0.1 | 87.8  0.5 |
| $Al_2O_3$ | Al | 25.7  2.57[a] | 29.93  1.12 | 30.35  0.05 | 29.68  0.1 | 21.1  0.2 |
| MgO | Mg | 30.4  3[c] | 19.8  0.4 | 19.85  0.1 | 20.1  0.15 | 19.9  0.2 |
| MgO+ | Mg | 63.9  6[c] | 47.5  0.7 | 47.4  0.2 | 48.0  0.28 | 19.8  0.16 |
| $Er_2O_3$ + | Er | 8.53  0.8[c] | 11  0.06 | 10.92  0.07 | 11.1  0.12 | 91.6  0.7 |
| $Er_2O_3$ | Er | 66.5  6[c] | 90.5  0.5 | 89.1  1.6 | 92.3  2 | 90.6  0.14 |
| MoN | Mo | 121  10[c] | 112.3  1.01 | 110.9  2.5 | NA | 27.9  0.4 |

[a] Value measured at ALS (STXM)
[b] Value measured at APS (SR-XRF)
[c] Value measured at ESRF (SR-XRF)
* part of the same multilayer
+ part of the same multilayer

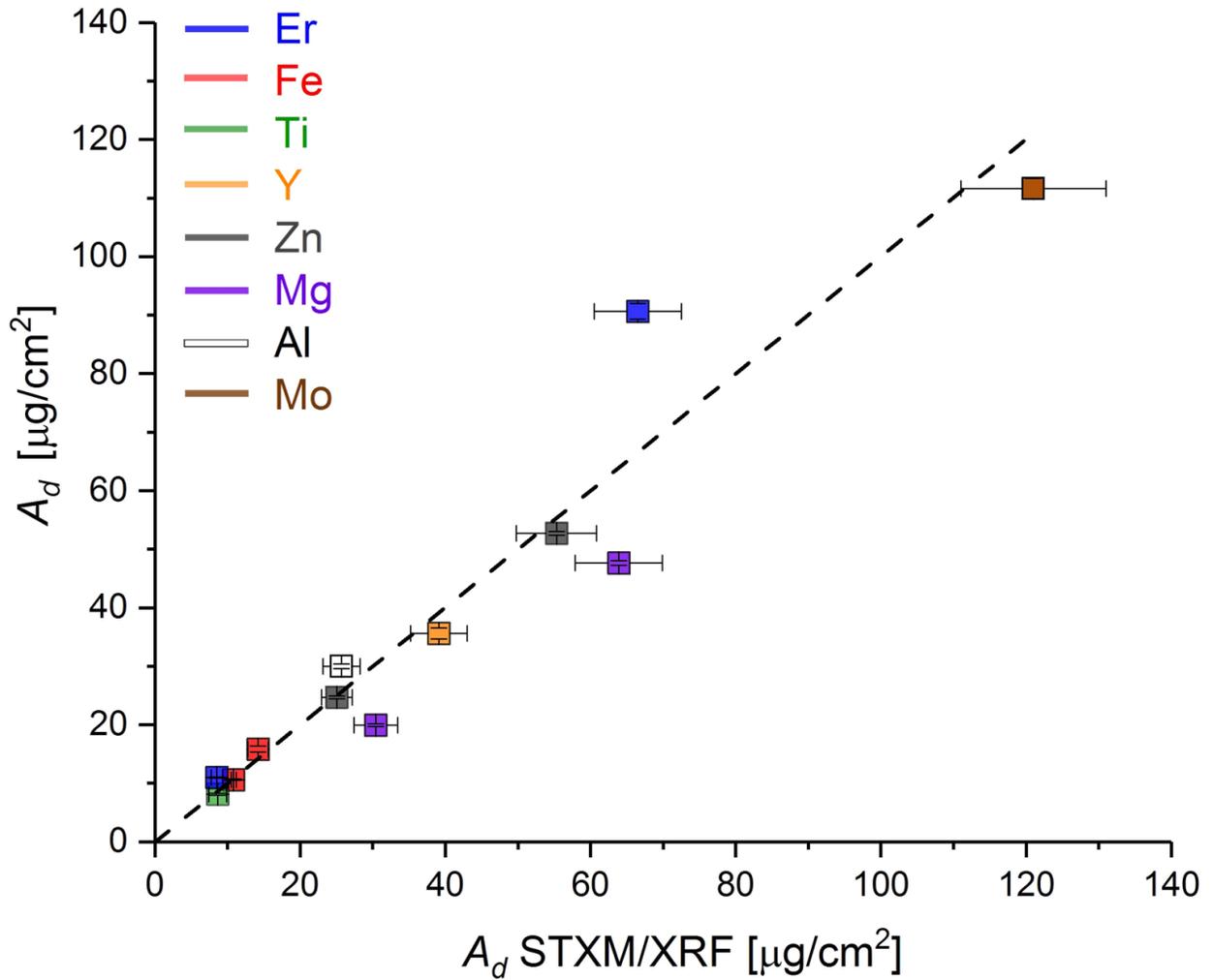

Figure 3: Summary of the data listed in Table 3. The dashed line represents a perfect correspondance between the STXM/XRF areal density measurement and the RBS, XRR, QCM averaged one.

of the TEM window. Taking for example the TMA and $H_2O$ process we can estimate the number of surface reactive sites using the same method as Elam, et. al.[37] to be $0.46 \times 10^{15}$ sites/cm$^2$, the thickness of the TEM window is 50 nm, H is 611.1 kJ/mol, and the density and specific heat of $Si_3N_4$ are 3.44 g/cm$^3$ and 800 J/(kg.K) respectively. Using equation 5 yields a ∆T of 80°C. While the TMA and water ALD process has a large self-limiting temperature range (up to 345°C) and can handle an increase in substrate temperature during growth many other processes, such as ZnO grown with DEZ and water, could be affected by such an increase. The ZnO process has an upper limit of 177°C, while the temperature increase of the $Si_3N_4$ membrane for a complete cycle, using equation 5, is ∆T 60‰. We overcame this parasitic CVD effect by either keeping the deposition temperature moderately low such that: $T_{dep} + ∆T < T_{Max}$. Where $T_{Max}$ is the upper limit of the ALD temperature range, and/or allowing sufficient cooling time by increasing the purge times after both half cycles. After these adjustments had been made the coatings were highly homogenous, as illustrated in figure 4.

As can be seen in figure 4 the use of perforated TEM windows is also viable. Depositing a conformal film on perforated windows allows for the measurement of the incident intensity, $I_0$, without the need to remove the standard and hence introduce possible beam fluctuations as a source of error. Sufficiently low temperature processes exist such that the deposition of ALD films on polymers has been achieved.[38,39] This points to the possibility of not only producing these standards on $Si_3N_4$ TEM windows, but on other materials that may be relevant in sample collection and preparation in the fields of biology and environmental science. This would allow for the direct measurement of absorption related to the substrate and alleviate the need to calculate these contributions in the measured specimens.

## 3D standards

The SEM pictures (Fig.5a-b)) of the ALD coated cubes shown in Fig.2c-d) reveal a homogeneous multilayer deposition; the new dimensions of the coated cube size and edges are 6.2 µm and 1.1 x 2.15 µm respectively.

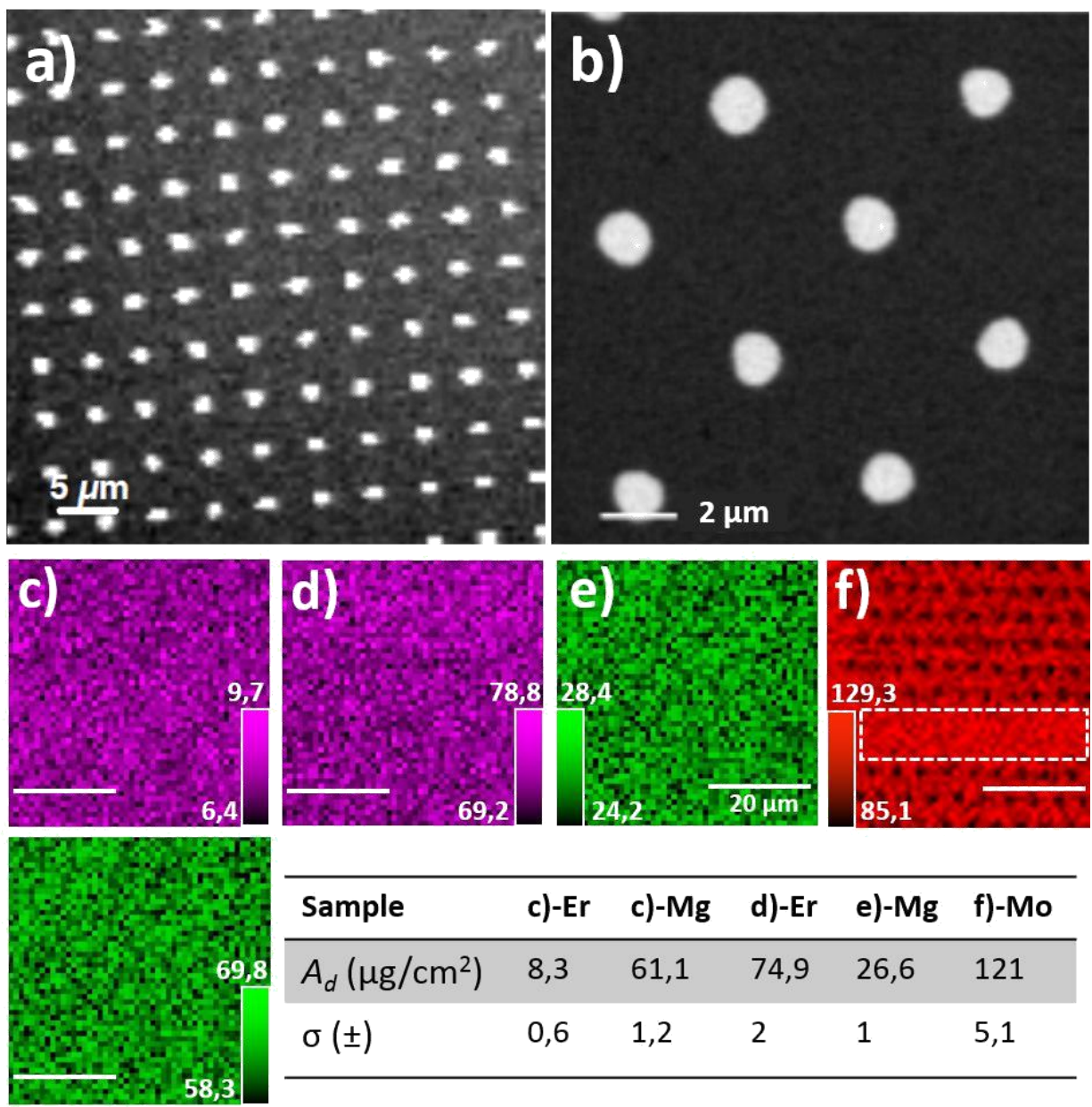

Figure 4: Homogeneity of the ALD films deposited on perforated and plain $Si_3N_4$ TEM windows and measured by STXM and -XRF. a) and b) STXM X-Ray absorption image of $Fe_2O_3$ coated windows acquired at a) 718 eV ALS beamline 11.0.2 and b) 740 eV at ALS beamline 5.3.2 showing homogeneity of the Fe distribution over wide area and nanometer scale. c-f) m-XRF cartographies of Er-L egde (magenta), Mg-K edge (green) and Mo-L edge (red) measured at ESRF beamline 21 at 2.7keV over 50x50 m$^2$ with 1 m steps. The scale bars are 20 m. The corresponding averaged values of the areal density $A_d$ and deviation are displayed in the table. For Mo, the averaged density value was taken from the plain area delimited by the dashed box in f).

Compared to bare cubes, each dimensions have been increased uniformly by 0.6 µm, which should coincide with twice the total deposited lm thickness. This analysis is consistent with the total lm thickness of 0.311 µm measured by XRR (Table2) on witness Si coupons. The TXM measurements and reconstruction of the coated cube (Fig.5b)) are shown in figure 5 c-f). The reconstructed cube side size is 6.4 µm and the edges dimensions are 1.2 0.1 µm 2.2 x 0.1 µm in very good agreement with the SEM pictures. The ALD multilayer coating appears as white and grey shells in figures 5c-e) with a total thickness of 0.35 x 0.05 µm in concordance with the SEM and XRR measurements. The outermost white layer corresponds to the thickest and highest Z ALD layer: ZnO with an estimated thickness of 0.1 m on the edge of this TXM measure resolution, but nonetheless consistent with XRR measurements. These encouraging results seed future work that will involve improving on the TXM resolution in order to spatially resolve the other ZnO layers of 4, 10, 20 and 40 nm deposited on the cube. Such 3D standards can also be used to test or develop reconstruction algorithms. In addition, the ZnO layers are polycrystalline as deposited on amorphous substrates with an average grain size, d, measured by XRD that depends on the lm thickness. According to ref.[40] for 50 nm thick lm d = 6 nm and for 125 nm d = 16 nm with what appears to be a saturation at d 20 nm for thicker films. This crystalline nature of ZnO in contrast with the as-grown amorphous $Al_2O_3$ could also be used to test the resolution limit of 3D reconstruction of nano X-ray diffraction techniques such as diffraction tomography.

It is noteworthy to mention that the polymer nature of the cube template can restrict the ALD temperature range and hence limit the choice of compounds that can be grown. For this reason, we purposefully chose alloys that can be synthesized at low temperature (165°C). Previous work[41] however have shown that pyrolysis of 3D printed structures under vacuum or inert atmosphere at temperature between 1000 to 3000°C can preserve the overall 3D printed shape while reducing significantly (up to 80%) the dimensions and transforming the polymer into glassy carbons, providing stable templates for any desired ALD alloys and growth temperature.

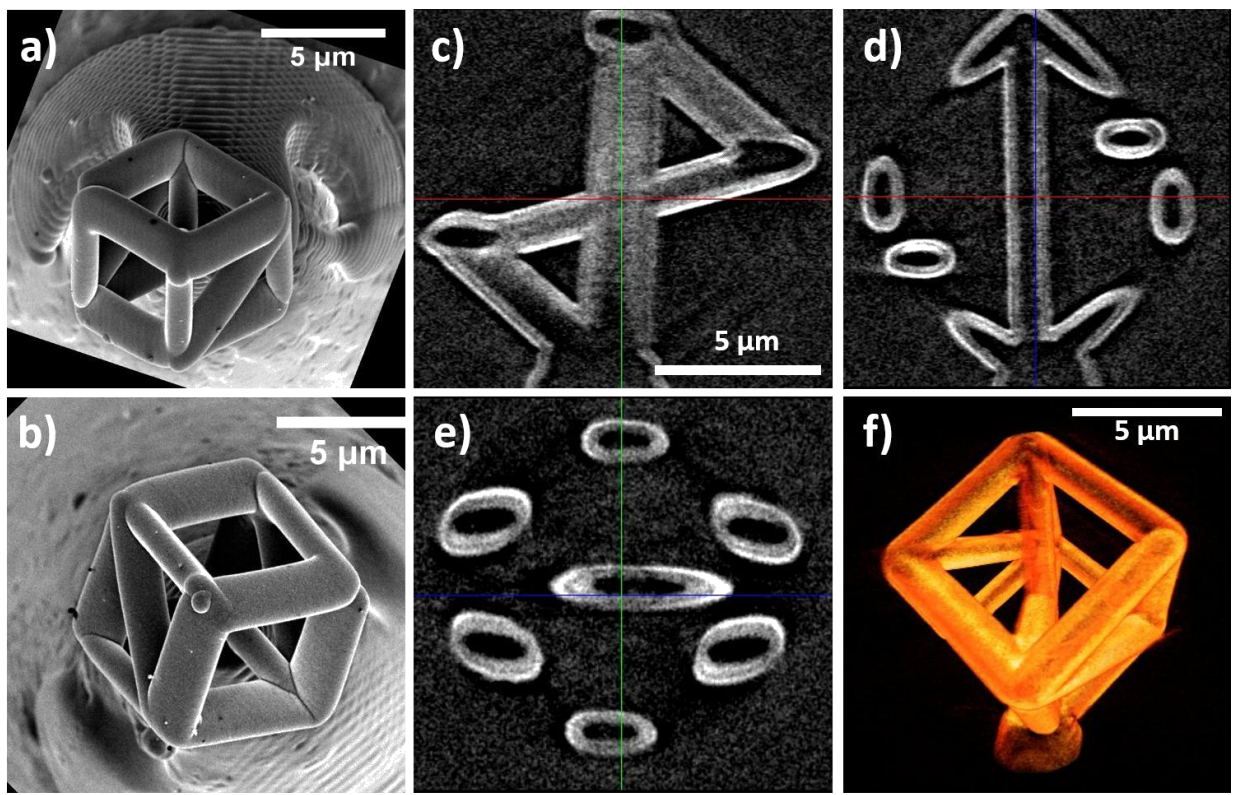

Figure 5: a) and b) SEM pictures of the same cubes shown in Fig.2 c) and d) after ALD coating with the Al2O3 and ZnO multilayer structure. The images have been rotated to match the orientations of Fig.2 c-d). c-f) 3D reconstruction of the ALD coated cube shown in b) and measured at APS beamline 32-ID-C at 8keV. c-d) are cuts of the reconstructed cube (f) along specific directions. The visible white lines correspond to the outermost ALD ZnO layer with a estimated thickness of 100 nm.

## Conclusion

The production of reliable SR-XRF/STXM standards, either multilayer or single compound films, produced via Atomic Layer Deposition has been demonstrated. Utilizing known chemistries we are able to produce standards for the majority of elements throughout of the periodic table thus eliminating the need to use NIST SRM 1832/1833 standards. The conformity and uniformity of ALD produced thin films has been demonstrated using a combination of X-ray characterization, SR-XRF, STXM, XRR, RBS and QCM. We have produced new custom-made TEM grid sample holder that enables the simultaneous coating and fabrication of up to 50 standards, in a timely and inexpensive manner. Moreover we have shown that the combination of ALD and 3D printing techniques provide suitable 3D standards with controlled shape and composition that can be extended to a large variety of structures. Along those lines, and as the 3D printing resolution improve with time, such synergy can be extended to the manufacturing of X-ray optics such as Fresnel zone plates, kinoform lenses, capillaries etc...

## Acknowledgment

This work was funded by the U.S. Department of Energy, Office of Sciences, Office of High Energy Physics, early Career Award FWP 50335 and Office of Science, Office of Basic Energy Sciences, Office of Science User Facility operated for the DOE Office of Science by Argonne National Laboratory under Contract No. DE-AC02-06CH11357. We also acknowledge the European Synchrotron Radiation Facility for provision of synchrotron radiation facilities at beamline 21, as well as resources of the Advanced Light Source, which is a DOE Office of Science User Facility under contract no. DE-AC02-05CH11231.

# References

(1) Benzerara, K.; Menguy, N.; Banerjee, N.; Tyliszczak, T.; Jr., G. B.; Guyot, F. Earth and Planetary Science Letters 2007, 260, 187-200.

(2) Obst, M.; Dynes, J.; Lawrence, J.; Swerhone, G.; Benzerara, K.; Karunakaran, C.; Kaz-natcheev, K.; Tyliszczak, T.; Hitchcock, A. Geochimica et Cosmochimica Acta 2009, 73, 4180-4198.

(3) Pecher, K.; McCubbery, D.; Kneedler, E.; Rothe, J.; Bargar, J.; Meigs, G.; Cox, L.; Nealson, K.; Tonner, B. Geochimica et Cosmochimica Acta 2003, 67, 1089-1098.

(4) Hub, C.; Burkhardt, M.; Halik, M.; Tzvetkov, G.; Fink, R. J. Mater. Chem. 2010, 20, 4884-4887.

(5) Yang, L.; McRae, R.; Henary, M. M.; Patel, R.; Lai, B.; Vogt, S.; Fahrni, C. J. Proceedings of the National Academy of Sciences of the United States of America 2005, 102, 11179-11184.

(6) Dik, J.; Janssens, K.; Van Der Snickt, G.; van der Loe, L.; Rickers, K.; Cotte, M. Analytical Chemistry 2008, 80, 6436-6442, PMID: 18662021.

(7) Vila-Comamala, J.; Dierolf, M.; Kewish, C. M.; Thibault, P.; Pilvi, T.; Farm, E.; Guzenko, V.; Gorelick, S.; Menzel, A.; Bunk, O.; Ritala, M.; Pfei er, F.; David, C. AIP Conference Proceedings 2010, 1221, 80-84.

(8) Twining, B. S.; Baines, S. B.; Fisher, N. S.; Maser, J.; Vogt, S.; Jacobsen, C.; Tovar-Sanchez, A.; Sanudo-Wilhelmy, S. A. Analytical Chemistry 2003, 75, 3806-3816.

(9) Punshon, T.; Guerinot, M. L.; Lanzirotti, A. Annals of Botany 2009, 103, 665{672.

(10) Moore, K. L.; Chen, Y.; van de Meene, A. M. L.; Hughes, L.; Liu, W.; Geraki, T.; Mosselmans, F.; McGrath, S. P.; Grovenor, C.; Zhao, F.-J. New Phytologist 2013,

(11) Ritala, M.; Niinisto, J. In Chemical Vapour Deposition: Precursors; Jones, A. C., Hitchman, M. L., Eds.; The Royal Society of Chemistry, 2009; pp 158-206.

(12) Ritala, M.; Leskel•a, M. In Handbook of Thin Films; Nalwa, H. S., Ed.; Academic Press: Burlington, 2002; pp 103-159.

(13) Puurunen, R. L. Journal of Applied Physics 2005, 97, 121301.

(14) Miikkulainen, V.; Leskel•a, M.; Ritala, M.; Puurunen, R. L. Journal of Applied Physics 2013, 113, 021301.

(15) Proslier, T.; Klug, J. A.; Elam, J. W.; Claus, H.; Becker, N. G.; Pellin, M. J. The Journal of Physical Chemistry C 2011, 115, 9477-9485.

(16) Klug, J. A.; Proslier, T.; Elam, J. W.; Cook, R. E.; Hiller, J. M.; Claus, H.; Becker, N. G.; Pellin, M. J. The Journal of Physical Chemistry C 2011, 115, 25063-25071.

(17) Riha, S. C.; Klahr, B. M.; Tyo, E. C.; Seifert, S.; Vajda, S.; Pellin, M. J.; Hamann, T. W.; Martinson, A. B. F. ACS Nano 2013, 7, 2396-2405.

(18) Hamann, T. W.; Martinson, A. B. F.; Elam, J. W.; Pellin, M. J.; Hupp, J. T. The Journal of Physical Chemistry C 2008, 112, 10303-10307.

(19) George, S. M. Chemical Reviews 2009, 110, 111-131.

(20) Elam, J.; George, S. Chemistry of Materials 2003, 15, 1020-1028.

(21) Lu, J.; Fu, B.; Kung, M. C.; Xiao, G.; Elam, J. W.; Kung, H. H.; Stair, P. C. Science 2012, 335, 1205-1208.

(22) Niinist•o, J.; Putkonen, M.; Niinist•o, L. Chemistry of Materials 2004, 16, 2953-2958.

(23) Klug, J. A.; Becker, N. G.; Riha, S. C.; Martinson, A. B. F.; Elam, J. W.; Pellin, M. J.; Proslier, T. J. Mater. Chem. A 2013, 1, 11607-11613.

(24) Aarik, J.; Karlis, J.; M•a, H., ndar; Uustare, T.; Sammelselg, V. Applied Surface Science 2001, 181, 339-348.

(25) Putkonen, M.; Sajavaara, T.; Niinist•o, L. Journal of Materials Chemistry 2000, 10, 1857-1861.

(26) P•aiv•asaari, J.; Niinist•o, J.; Arstila, K.; Putkonen, M.; Niinist•o, L. Chemical Vapor Deposition 2005, 11, 415-419.

(27) Klug, J. A.; Becker, N. G.; Groll, N. R.; Cao, C.; Weimer, M. S.; Pellin, M. J.; Zasadzinski, J. F.; Proslier, T. Journal of Materials Chemistry 2013, 103, 211602.

(28) Henke, B.; Gullikson, E.; Davis., J. Atomic Data and Nuclear Data Tables 1993, 54, 181-342.

(29) De Andrade, V.; Deriy, A.; Wojcik, M.; Grsoy, D.; Shu, D. I.; Fezzaa, K.; De Carlo, F. SPIE Newsroom 2016,

(30) Grsoy, D.; De Carlo, F.; Xiao, X.; Jacobsen, C. Journal of Synchrotron Radiation 2014, 21, 1188-1193.

(31) De Carlo, F.; Grsoy, D.; Marone, F.; Rivers, M.; Parkinson, Y.; Khan, N., F. Schwarz; D.J., V.; Vogt, S.; S.C., G.; Narayanan, S.; Newville, T., M. Lanzirotti; Sun, Y.; Hong, Y.; Jacobsen, C. Journal of Synchrotron Radiation 2014, 21, 1224-1230.

(32) Miqueles, E.; Rinkel, J.; O'Dowd, F.; Bermudez, J. Journal of Synchrotron Radiation 2014, 21, 1333-1346.

(33) Pelt, D.; Batenburg, K. Proceedings of the 2015 International Meeting on Fully Three-Dimensional Image Rreconstruction in Radiology and Nuclear Medicine 2014, 21, 1333-1346.

(34) Pelt, D.; Grsoy, D.; Palenstijn, W.; Sijbers, J.; De Carlo, F.; Batenburg, K. Journal of Synchrotron Radiation 2016, 23, 842-849.

(35) Pelt, D.; De Andrade, V. Advanced Structural and Chemical Imaging 2016, 2, 17.

(36) Liu, G.; Deguns, E.; Lecordier, L.; Sundaram, G.; Becker, J. ECS Transactions 2011, 41, 219-225.

(37) Elam, J. W.; Routkevitch, D.; Mardilovich, P. P.; George, S. M. Chemistry of Materials 2003, 15, 3507-3517.

(38) Elam, J. W.; Wilson, C. A.; Schuisky, M.; Sechrist, Z. A.; George, S. M. Journal of vacuum science & technology. B, Microelectronics processing and phenomena 2003, 21, 1099.

(39) Minton, T. K.; Wu, B.; Zhang, J.; Lindholm, N. F.; Abdulagatov, A. I.; O'Patchen, J.; George, S. M.; Groner, M. D. ACS Applied Materials & Interfaces 2010, 2, 2515-2520.

(40) Chaaya, A. A.; Viter, R.; Bechelany, M.; Alute, Z.; Erts, D.; Zalesskaya, A.; Kovalevskis, K.; Rouessac, V.; Smyntyna, V.; Miele, P. Beilstein Journal of Nanotechnol-ogy 2013, 4, 690-698.

(41) Bauer, J.; Schroer, A.; Schwaiger, R.; Kraft, O. nature materials 2013, 4, 438-444.